\documentclass[]{spie}  %>>> use for US letter paper
\usepackage[]{graphicx}
\usepackage{epstopdf}

\title{Collision-free motion planning for fiber positioner robots: discretization of velocity profiles} 

\author{Laleh Makarem\supit{a}, Jean-Paul Kneib\supit{b,c}, Denis Gillet\supit{a}, Hannes Bleuler\supit{d}, Mohamed Bouri\supit{d}, Philipp H\"{o}rler\supit{d}, Laurent Jenni\supit{d},  Francisco Prada\supit{e,f} and Justo Sanchez\supit{e}
\skiplinehalf
\supit{a}Coordination and Interaction Systems Group (REACT), Ecole Polytechnique F\'{e}d\'{e}rale de Lausanne (EPFL), Switzerland \\
\supit{b}Laboratory of Astrophysics (LASTRO), Ecole Polytechnique F\'{e}d\'{e}rale de Lausanne (EPFL), Observatoire de Sauverny, Ch-1290 Versoix, Switzerland\\
\supit{c}Aix Marseille Universit\'{e}, CNRS, LAM (Laboratoire d'Astrophysique de Marseille) UMR 7326, 13388, Marseille, France\\
\supit{d}Laboratory of Robotic Systems (LSRO)Ecole Polytechnique F\'{e}d\'{e}rale de Lausanne (EPFL), Switzerland\\
\supit{e}Instituto de Astrofisica de Andalucia (CSIC), Granada, E-18008, Spain\\
\supit{f}Instituto de F\'{i}sica Te\'{o}rica, (UAM/CSIC), Universidad Aut\'{o}noma de Madrid, Cantoblanco, E-28049 Madrid, Spain
}

\authorinfo{Corresponding author: laleh.makarem@epfl.ch}
\begin{document} 
  \maketitle 

%%%%%%%%%%%%%%%%%%%%%%%%%%%%%%%%%%%%%%%%%%%%%%%%%%%%%%%%%%%%% 
\begin{abstract}
The next generation of large-scale spectroscopic survey experiments such as DESI, will use thousands of fiber positioner robots packed on a focal plate. In order to maximize the observing time with this robotic system we need to move in parallel the fiber-ends of all positioners from the previous to the next target coordinates. Direct trajectories are not feasible due to collision risks that could undeniably damage the robots and impact the survey operation and performance. We have previously developed a motion planning method based on a novel decentralized navigation function for collision-free coordination of fiber positioners. The navigation function takes into account the configuration of positioners as well as their envelope constraints. The motion planning scheme has linear complexity and short motion duration (~2.5 seconds with the maximum speed of 30 rpm for the positioner), which is independent of the number of positioners. These two key advantages of the decentralization designate the method as a promising solution for the collision-free motion-planning problem in the next-generation of fiber-fed spectrographs. In a framework where a centralized computer communicates with the positioner robots,  communication overhead can be reduced significantly by using velocity profiles consisting of a few bits only. We present here the discretization of velocity profiles to ensure the feasibility of a real-time coordination for a large number of positioners. The modified motion planning method that generates piecewise linearized position profiles guarantees collision-free trajectories for all the robots. The velocity profiles fit few bits at the expense of higher computational costs.  
\end{abstract}

%>>>> Include a list of keywords after the abstract 


%%%%%%%%%%%%%%%%%%%%%%%%%%%%%%%%%%%%%%%%%%%%%%%%%%%%%%%%%%%%%
\section{INTRODUCTION}
\label{sec:intro}  % \label{} allows reference to this section

Massive spectroscopic surveys allow cosmologists not only to measure the redshifts of distant objects, but also to trace the large-scale structure of our Universe. CfA, the first massive spectroscopic survey of the local Universe, revealed a cosmic web structure with filaments and voids \cite{huchra1983survey}. CfA was followed by many significant cosmological surveys such as sloan digital sky survey (SDSS) \cite{york2000sloan} and the 2-degree field galaxy redshift survey (2dFGRS) \cite{colless20012df}. The two surveys independently measured the signature of the baryonic acoustic oscillations (BAO). The specific shape of the signature of BAO in the power-spectrum candidates BAO as one of the least biased probes to trace the expansion history of the Universe.

In light of promising characteristics of BAO, Blake and his colleagues completed the WiggleZ,  a $\sim 250,000$ redshift survey of star-forming galaxies (at $z < 0.8$), which was performed on the 4m Anglo australian telescope.  Succeeding the WiggleZ, baryonic oscillation spectroscopic survey (BOSS) 
will complete a major redshift survey of 1.4 million galaxy redshift quasars (at $z < 0.7$) and 160,000 high-redshift Lyman-quasars using the SDSS telescope \cite{anderson2012clustering}. The extended-BOSS, eBOSS, which has recently been started will complete the first BAO survey of galaxies and quasars over the redshift range of $0.7 < z < 2.2$ in six years using the SDSS facility.

\begin{figure}
   \begin{center}
   \begin{tabular}{c}
   \includegraphics[width=0.7\textwidth]{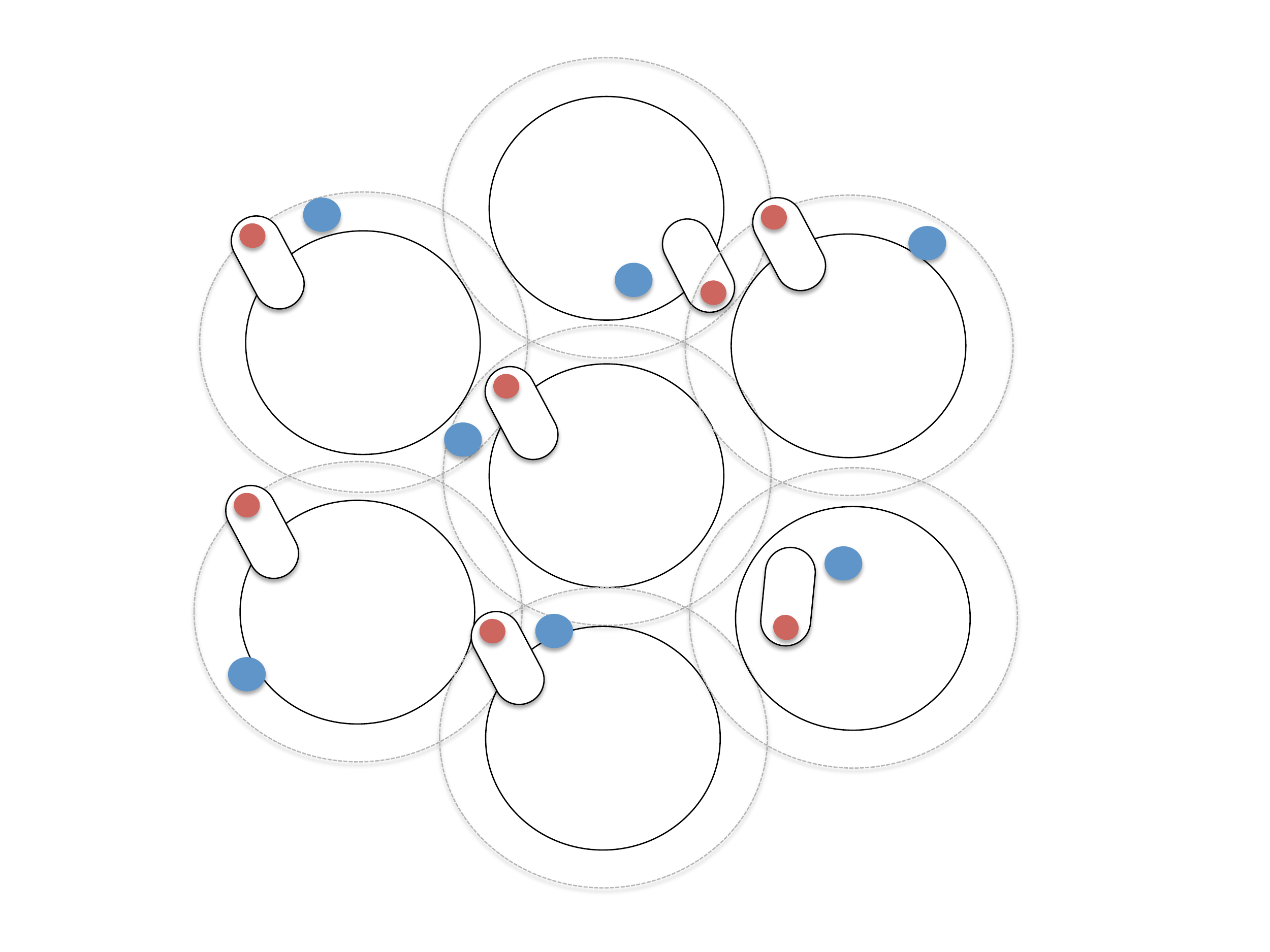}
   \end{tabular}
   \end{center}
   \caption{ \label{Fig:WorkspaceSharing} 
   Positioner robots will be packed closely in a focal plate. To ensure their access to all part of the focal plate, robots should share workspace. Red and blue dots, respectively  show the fiber ends and target points.}
   \end{figure}

\begin{figure}
   \begin{center}
   \begin{tabular}{c}
   \includegraphics[width=0.4\textwidth]{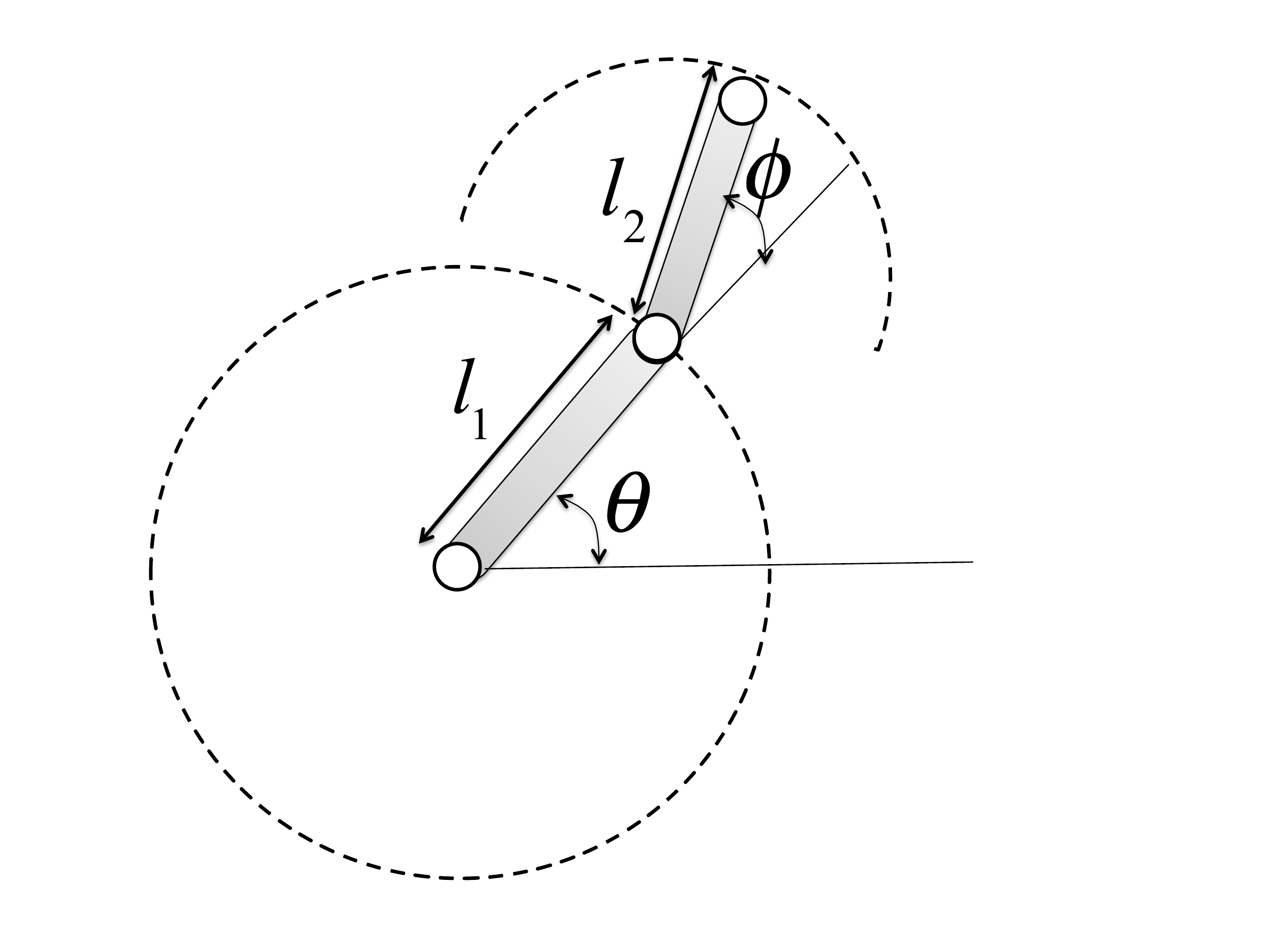}
   \end{tabular}
   \end{center}
   \caption{ \label{Fig:2DOF_accentric} 
   A two-degree of freedom robot with rotary joints. A robot with such kinematics could also be called $\theta - \phi$ design in astronomical instrumentation.}
   \end{figure}

By developing massive parallel redshift measurement instruments we can make a paradigm shift in the observation techniques compared with what is currently possible in BOSS and eBOSS. In the next-generation fiber-fed spectrographs such as DESI, small robots will position the fiber ends simultaneously. In order to improve the versatility of the system and to ensure the maximum number of observed galaxies, the robots share workspaces in a focal plate (Figure~\ref{Fig:WorkspaceSharing}). Sharing the workspace increases the chance of collision among the adjacent robots. In our previous work \cite{makarem2013collision}, we tackled this problem for a robot with two degrees of freedom and eccentric rotary joints the so-called $\theta - \phi$ design (Figure \ref{Fig:2DOF_accentric}). The proposed method could directly overcome the collision problem for any robot with two rotary joints that move a single fiber end such as for the DESI,  PFS, or MOONS projects. The proposed motion planning method is based on decentralized navigation functions (DNF). The proposed trajectories guarantee collision-free path for all the fiber ends. The motion planning is decentralized in order to enable the extension of  the solution for larger-scale positioner robots.
%%%%%%%%%%

One of the main candidates among communication solutions  for DESI project is a framework based on wired communication. According to this framework, all placement orders, configuration messages, and alert advices will be sent from a centralized master computer to each actuator through a fast plus $I^2C$ bus, by which we can achieve a transfer rate up to 3.4Mbps.   In order to minimize the communication time the profiles of motor velocities or positions should be discretized to fit few bits. Moreover, by discretization of positioner profiles we can use a standard electronics and mostly focus on thermal noise minimization in the control level. So, we present here the pieces-wise linearization of position profiles to ensure the feasibility of a real-time coordination for a large number of positioners. The modified motion planning method guarantees collision-free paths for all the robots in expense of more off-line computational costs.  

This paper is organized as follows. In Section 2 we briefly give the problem formulation and describe how collision-free trajectories are calculated. The piece-wise linearization algorithm used to discretize the profiles is explained in Section 3.  The simulation results corresponding to the proposed approach are presented and discussed in Section 4. Finally, in Section 5, we explore the possibilities for future research and conclude. 

\section{Collision-free motion planning}

For acquiring collision-free trajectories for all the 5000 positioners, we assume that target assignment has been effectively done under three main assumptions. First, each target is assigned to one positioner only, and each positioner is assigned to maximum one target. Second, the target assigned to each positioner is always within its patrol disc and hence reachable by the positioner.  The distance between two assigned targets is not less than $D$. So,  we assume for this step that the target of each positioner is fixed and known to the positioner robot. Thus, the focus of the work presented here is on the coordination of positioners in motion to avoid collision. Moreover, we assume that the target of each positioner is fixed and known to the positioner robot.
%%%%%%%%%%%%%%%%%%%%%%%%%%%%%%%%%%%%%%%%%%%%%%%%%%%%%%%%%%%%%
\subsection{Configuration of the positioners}

We consider a system composed of $N$ positioner robots. The goal of each robot is to put its end effector (fiber-end) on an assigned target point.
Each positioner covers the patrol area (workspace) through two correlated rotations, $\theta$ and $\phi$ (Fig.~\ref{Fig:2DOF_accentric})

The forward kinematics of each positioner robot can be described by:
\begin{equation}
\label{DynamicModel}
\left( {\begin{array}{*{20}c}
   {x_i }  \\
   {y_i }  \\
\end{array}} \right) = \left( {\begin{array}{*{20}c}
   {x_{ib} }  \\
   {y_{ib} }  \\
\end{array}} \right) + \left( {\begin{array}{*{20}c}
   {cos\theta_{i} } & {cos(\theta_{i}  + \phi_{i} )}  \\
   {sin\theta_{i} } & {sin(\theta_{i}  + \phi_{i} )}  \\
\end{array}} \right)\left( {\begin{array}{*{20}c}
   {l_1 }  \\
   {l_2 }  \\
\end{array}} \right)
\end{equation}

\begin{figure}
   \begin{center}
   \begin{tabular}{c}
   \includegraphics[width=0.9\textwidth]{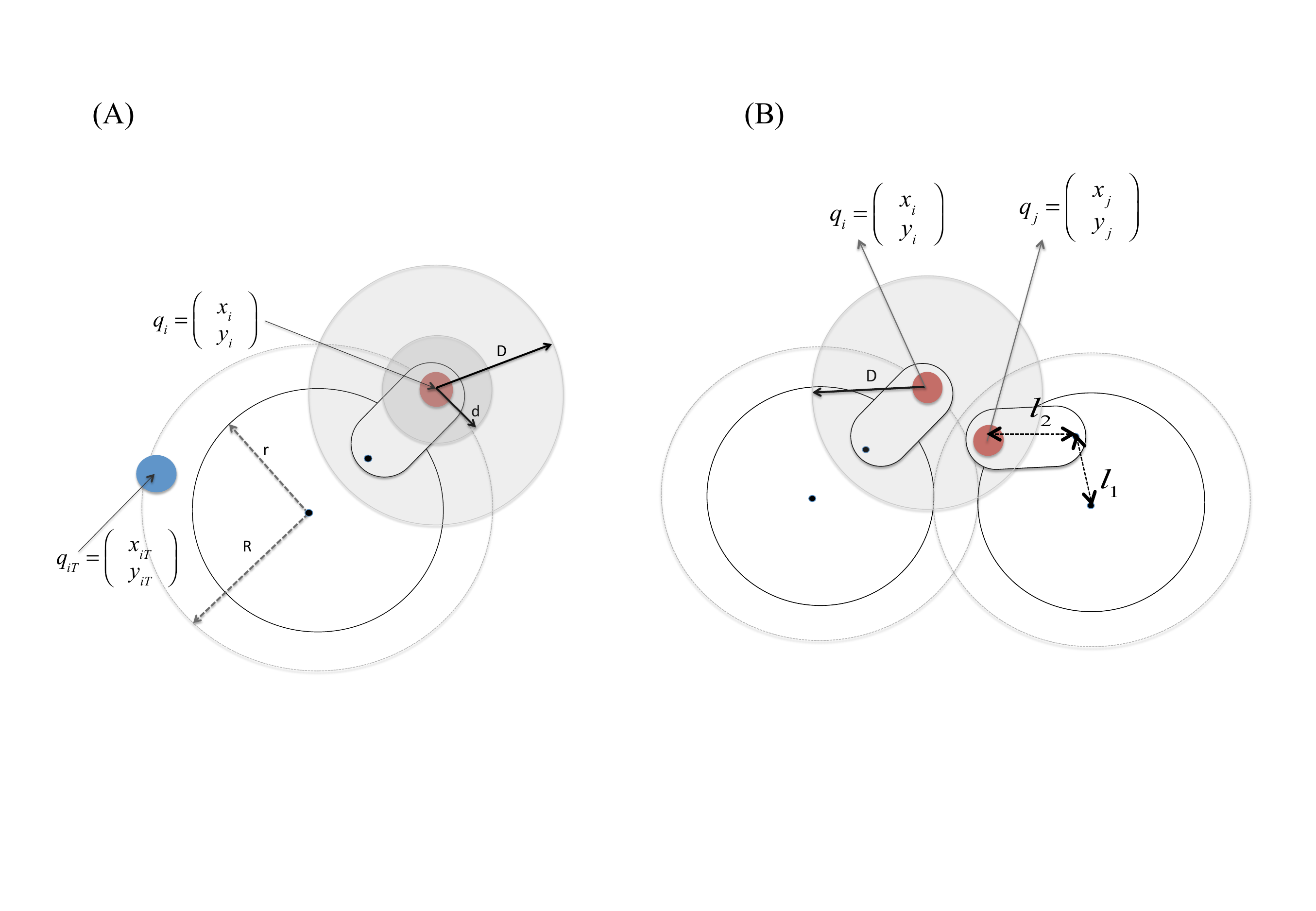}
   \end{tabular}
   \end{center}
   \caption{ \label{Fig:Schematics} 
(A) The target point $q_{iT}$ is the destination of the robot end-effector. The collision detection envelope with radius of D, is the area where collision avoidance term in the navigation function is activated. The collision avoidance term in the navigation function ensures that the two positioner end-effectors will keep a distance larger than $d$. (B) A configuration in which there is a risk of collision between the end-effectors of two robots. In this configuration the collision avoidance term in the navigation functions of the two robots are active which means they take values more than zero. }
   \end{figure}

Where the end-effector position of positioner $i$ is $q_i = (x_i,y_i)$  in a global frame attached to the focal plane. $l_1$ and $l_2$ are first and second rotation links respectively (Fig.~\ref{Fig:Schematics}).  $\theta$ and $\phi$ are angular positions of the two joints of the positioner $i$. Each positioner is controllable by its angular velocity, meaning the speed of each of the two motors.
   
The main challenge is to coordinate the robots in motion to reach the predefined targets while avoiding collisions. The proposed approach should be expandable to a more large-scale problems. A centralized solution for such a problem would be practically infeasible and computationally costly due to the presence of numerous positioners and constraints (\cite{tanner2005towards}). Therefore, among all the methods found in the literature for coordinating agents, decentralized and reactive control approaches are more efficient.
%%%%%%%%%%%%%%%%%%%%%%%%%%%%%%%%%%%%%%%%%%%%%%%%%%%%%%%%%%%%%
 \subsection{Decentralized navigation function}

Inspired by the emergent behaviors among swarms (insects, birds, fishes), methods based on local reactive control have received great interest (\cite{ge2012decentralized}). Therefore, artificial potential fields are often exploited for the coordination of autonomous agents. The main drawback of most potential field approaches is that convergence to the target is not guaranteed due to the presence of spurious local minima. In order to solve this problem and present a complete and exact solution for the coordination problem, navigation functions have been introduced (\cite{dimarogonas2005decentralized}).

Navigation functions are used in various robotic and control applications (\cite{makarem2011decentralized, makarem2012fluent, de2006formation}). In these applications the actuation torque or other inputs (e.g., the acceleration, the velocity) is derived from some potential function that encodes relevant information about the environment and the objective. In the framework of the problem presented in this paper, the use of navigation functions in a decentralized scheme is promising, as it can be implemented in real-time and it also shows good flexibility with regard to adding new robots and changing the problem constraints.

A navigation function is an analytic function from the workspace of every positioner robot and its gradient would be attractive to its destination and repulsive from other robots. So, an appropriate navigation function could be combined with a proper control law in order to obtain a trajectory for every motor of the robot leading to the destination and avoiding collisions.

{We propose a navigation function $\psi_i$ for the positioner $i$ in (\ref{DNF}) which is composed of two components. The first term, the attractive component, is the squared distance of the end-effector of positioner robot $i$ from its target point. This term of navigation function attains small values as the positioner brings the fiber closer to its target point (Fig. \ref{Fig:Schematics} (A)). The second term, the repulsive component, aims at avoiding collisions between positioner $i$ and the six other positioners located in its vicinity. This term is activated when the two positioner robots are closer than a distance $D$, otherwise this term is zero. $D$ defines the radius of a collision avoidance envelope and $d<D$ defines the radius of the safety region. The closer the two robots get, the higher values this repulsive term attains. Moving toward the minimum point of this navigation function will guarantee the minimum distance of $d $ between the positioners (Fig. \ref{Fig:Schematics} (B)). $\lambda_1$ and $\lambda_2$ are the two weighting parameters related to the two terms in the navigation function.}

\begin{eqnarray}
\psi _i  = \lambda_1 \left\| {q_i  - q_{iT} } \right\|^2 + \lambda_2 \sum\limits_{j \ne i} {min(0,\frac{{\left\| {q_i  - q_j } \right\|^2  - D^2 }}{{\left\| {q_i  - q_j } \right\|^2  - d^2 }})} 
\label{DNF}
\end{eqnarray}

According to the navigation function presented in (\ref{DNF}) and the forward kinematics defined in (\ref{DynamicModel}), the following control law is proposed:
\begin{equation}
\label{Control}
u_i  =  - \nabla _{\theta _i  \phi_i} \psi _i (q_i )
\end{equation}
At each step, the robot will move the fiber according to a gradient descent method. It is worth mentioning that, the navigation function is directly a function of the end-effector positions. In order to obtain the angular velocities of each of two motors, we calculate the gradient of the navigation function with respect to the angular positions of the links using the chain derivatives and the forward kinematics in (\ref{DynamicModel}).\\
\begin{equation}
\label{ChainDer}
\left( {\begin{array}{*{20}c}
   {\omega _{i1} }  \\
   {\omega _{i2} }  \\
\end{array}} \right) = u_i  =  - \left( {\begin{array}{*{20}c}
   {\frac{{\partial \psi _i }}{{\partial x_i }}\frac{{\partial x_i }}{{\partial \theta_{i} }} + \frac{{\partial \psi _i }}{{\partial y_i }}\frac{{\partial y_i }}{{\partial \theta_{i} }}}  \\
   {\frac{{\partial \psi _i }}{{\partial x_i }}\frac{{\partial x_i }}{{\partial \phi_{i} }} + \frac{{\partial \psi _i }}{{\partial y_i }}\frac{{\partial y_i }}{{\partial \phi_{i} }}}  \\
\end{array}} \right)
\end{equation}
$\omega_{i1}$ and $\omega_{i2}$ are the angular velocity of the first and second motor of the positioner robot $i$ respectively. There could be two final configurations that result in the same target point. However, when traveling from a known configuration, reaching to one of those configurations is faster considering no collision. In this work we assumed that the faster configuration is the goal configuration. 

In \cite{makarem2013collision}, we provided analytical  proofs for collision avoidance and convergence of the DNF introduced by \ref{DNF} for DESI specifications. In this paper our aim is to discretize the calculated actuator profiles, to facilitate sending profiles to the robots in practice .  
%%%%%%%%%%%%%%%%%%%%%%%%%%%%%%%%%%%%%%%%%%%%%%%%%%%%%%%%%%%%%

\section{Profile discretization}

\begin{figure}
   \begin{center}
   \begin{tabular}{c}
   \includegraphics[width=1\textwidth]{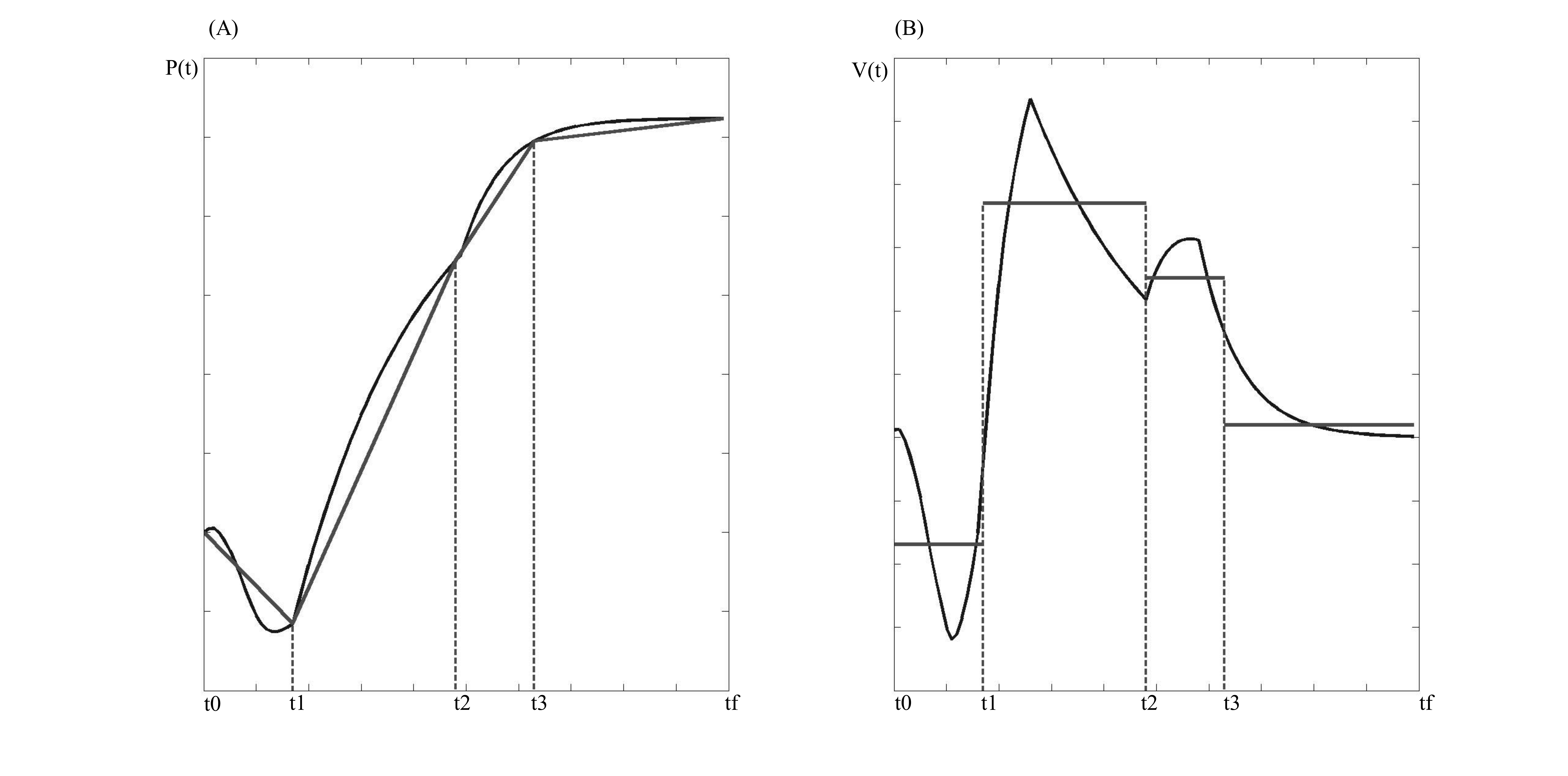}
   \end{tabular}
   \end{center}
   \caption{ \label{Fig:VelPosProfile}  (A) An example of position profile calculated using DNF for motion planning. (B) A velocity profile of a motor, corresponding to the position profile in the left. By piecewise linearizing the position profile, the velocity profile is piecewise constant.}
   \end{figure}

Consider  $v(t)$ and $p(t)$ general velocity and its corresponding position profile for a motor respectively (Fig.~\ref{Fig:VelPosProfile}).  $p(t)$ is a continuous nonlinear function, of a single variable $t$ in an interval $[t_0, t_f]$.  To send the minimum data to the motors the goal is to find minimum number of intervals of constant velocity and send the velocity and its duration to the motors. To this end, we approximate the position profiles by piecewise linear functions. Consequently, the velocity profiles will be piecewise constant functions. When approximating the position profiles, the collision avoidance constraints  should be considered. 
     
\begin{figure}
   \begin{center}
   \begin{tabular}{c}
   \includegraphics[width=1\textwidth]{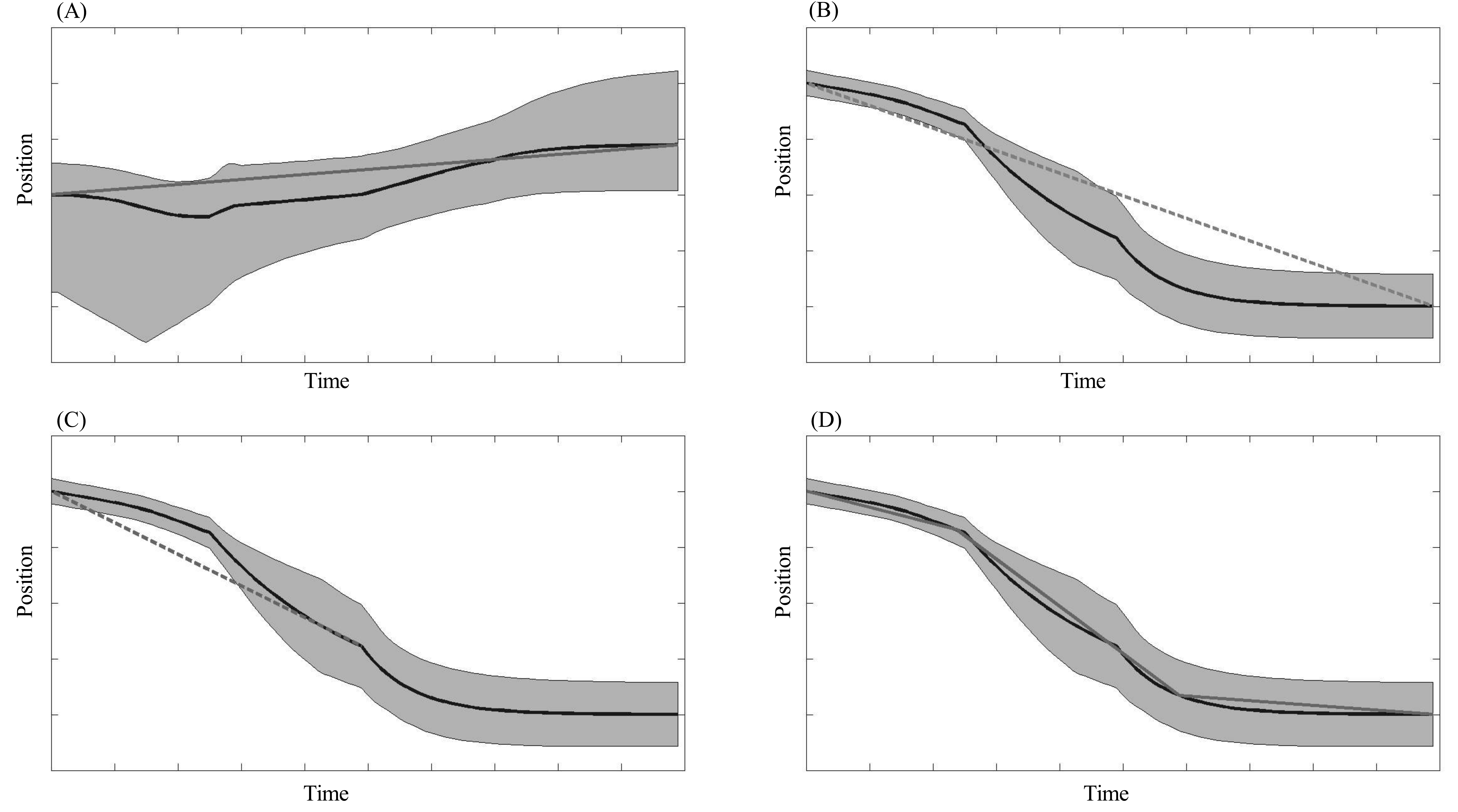}
   \end{tabular}
   \end{center}
   \caption{ \label{Fig:DisProfiles} (A) A position profile with the safety margin shown in grey. In the first step, we draw a line from $t_0$ to $t_f$ on the profile. This line is entirely in the safety zone. By changing the position profile from the continuous curve to the single line, still collision avoidance is guaranteed. (B,C,D) A single line can not provide a collision free trajectory for the positioner. (C)Therefore, we find a middle point between $t_0$ and $t_f$ as a breaking point. But the line is not still entirely in the safety margin, so we need to repeat the last step, till we get line segments entirely in the safety margin (D). }
\end{figure}

Piecewise linear functions are used in many applications and therefor various methods have been introduced in the literature. The main focus of piecewise linearization appears on optimization problem with a nonlinear cost function \cite{sontag1981nonlinear,rantzer2000piecewise,rewienski2003trajectory}. The main goal in many different applications is to find the number and location of the break points along the function. In our application, due to collision risk the breaking points play an important role. On the other hand, lower number of breaking points means less communication cost in the electronics side. Keeping these two constraints in mind, the position profiles are linearized by the following algorithm:\\
\textbf{Step one}: the position profile of the first motor is filled with a safety margin. The safety margin is the projection of the half the distance from the positioner's fiber end to that of its nearest adjacent positioner on the angular position of the motors.  (Fig.~ \ref{Fig:DisProfiles}).\\
\textbf{Step two}: draw a line from $t_0$ to $t_f$ on the position profile.\\
\textbf{Step three}: If the whole line is inside the safety margin, the profile has been linearized with only one line segment (Fig.~ \ref{Fig:DisProfiles} (A)). If the line passes over the safety margin put $t_n = t_0$ (Fig.~ \ref{Fig:DisProfiles} (B))\\
\textbf{Step four}: find the mean of the start point ($t_n$) and the final point ($t_f$), call it $t_m$.\\
\textbf{Step five}: draw a line from  $t_n$ to $t_m$  on the position profile (Fig.~ \ref{Fig:DisProfiles} (C)).\\
\textbf{Step six}: If the whole line is inside the safety margin, one part of the profile has been linearized with one line segment. put $t_n = t_m$ and repeat from step five, else find the mean value of the points $t_n$ and $t_m$ and put is as $t_m$ and repeat from step five (Fig.~ \ref{Fig:DisProfiles} (D)).\\
\textbf{Step seven}: compute the gradient of the line segments, these correspond to the velocity profiles.  

For the second motor, we should take into account the new profile of the first motor. The safety margin needs to be recalculated.
It is worth mentioning that, calculating the safety margin does not add a big computational load, since in each time step during the motion planning algorithm the distance of each two adjacent positioner is computed. We only need to store that data to use further for piecewise linearization of position profiles.
   
 %%%%%%%%%%%%%%%%%%%%%%%%%%%%%%%%%%%%%%%%%%%%%%%%%%%%%%%%%%%%%
\section{Simulations and results}

\begin{figure}[t]
   \begin{center}
   \includegraphics[width=1\textwidth]{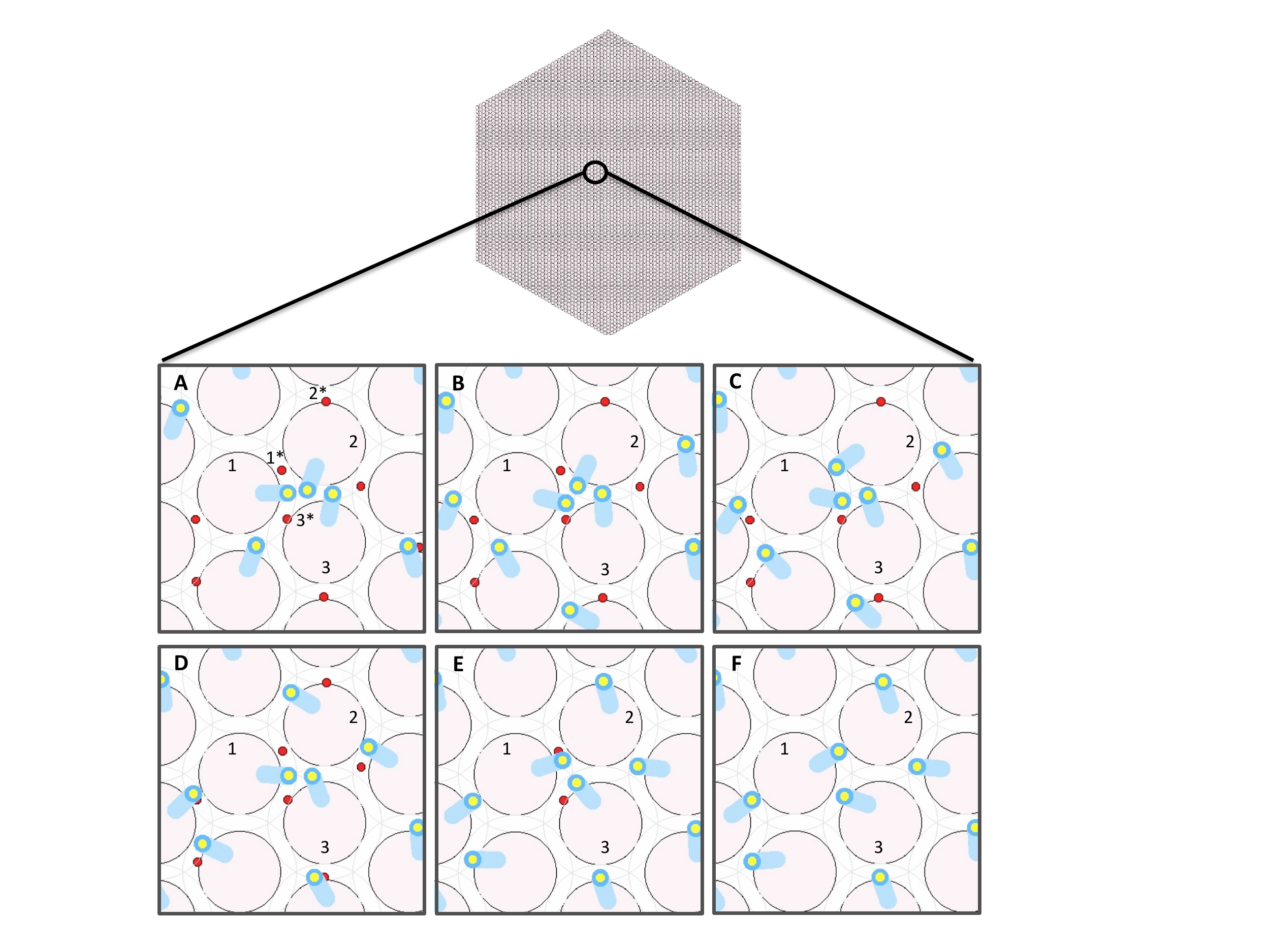}
   \end{center}
   \caption{ \label{Fig:Simulator} Simulation environment for 5000 positioners and a zoom on the motion of three positioners. The six boxes (A to F) show six snap shots of the simulation. 1*, 2* and 3* are the target positions for the three positioners 1,2 and 3. These three positioners are engaged in a local conflict in which they need specific maneuvers to get to their destination.}
   \end{figure}

We have developed a visual simulation environment (See Fig.~\ref{Fig:Simulator}) for the whole 5000 positioners for the test and validation of motion planning, collision avoidance, and profile discretization algorithms. In Fig.~\ref{Fig:Simulator}, the six boxes (A to F) show six snapshots of the simulation. Star points show the target positions for positioners 1,2, and 3. These snapshots are selected in a way that they show one of the most challenging scenarios where collision avoidance is very important. Positioner 1 needs to make space for positioner 2 to pass. Positioner 2 cannot make room for positioner 1 because positioner 3 is blocking the way. Positioner 3 needs to pass both positioners to reach its target point. The small maneuver from positioner 1 that comes directly from DNF moves this positioner farther from its target point, but it makes room for positioner 2 to pass safely. When positioner 2 clears the way, positioner 1 starts moving toward its target point and this gives a safe way to the positioner 3. The small maneuvers and small safety margin makes this scenario challenging for profile discretization as well. 

\begin{figure}
   \begin{center}
   \begin{tabular}{c}
   \includegraphics[width=1\textwidth]{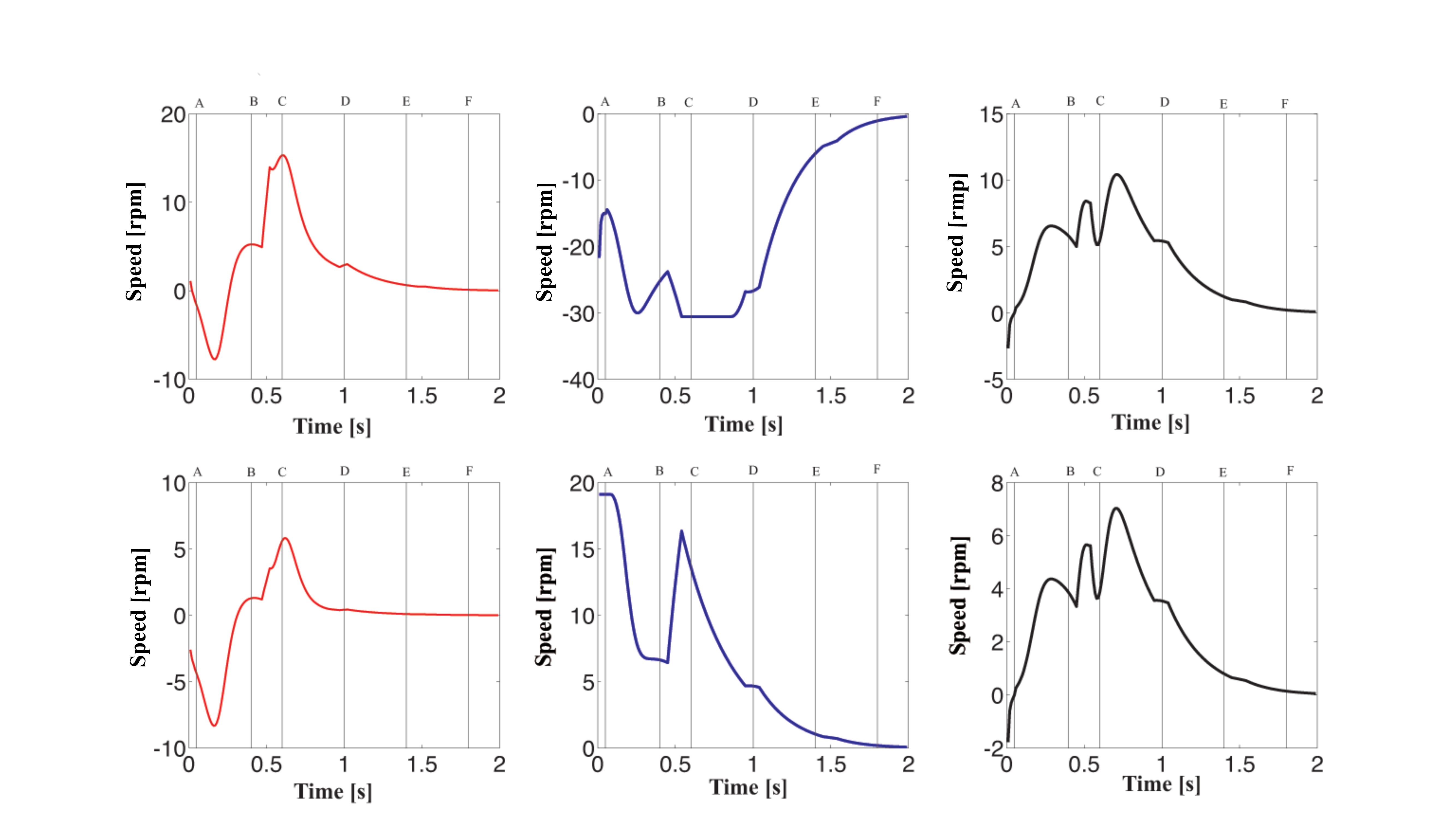}
   \end{tabular}
   \end{center}
   \caption{ \label{Fig:VelocityProfiles}Velocity profiles that correspond to the pairs of actuators for positioners 1,2, and 3 in Fig.~\ref{Fig:Simulator}. Columns show the velocity profiles for each positioner. The first and second profiles of each column correspond to the first and second actuator of each positioner, respectively. Vertical lines indicate the moment at which the snapshots in Fig.~\ref{Fig:Simulator}  were taken.  }
   \end{figure}

\begin{figure}
   \begin{center}
   \begin{tabular}{c}
   \includegraphics[width=1\textwidth]{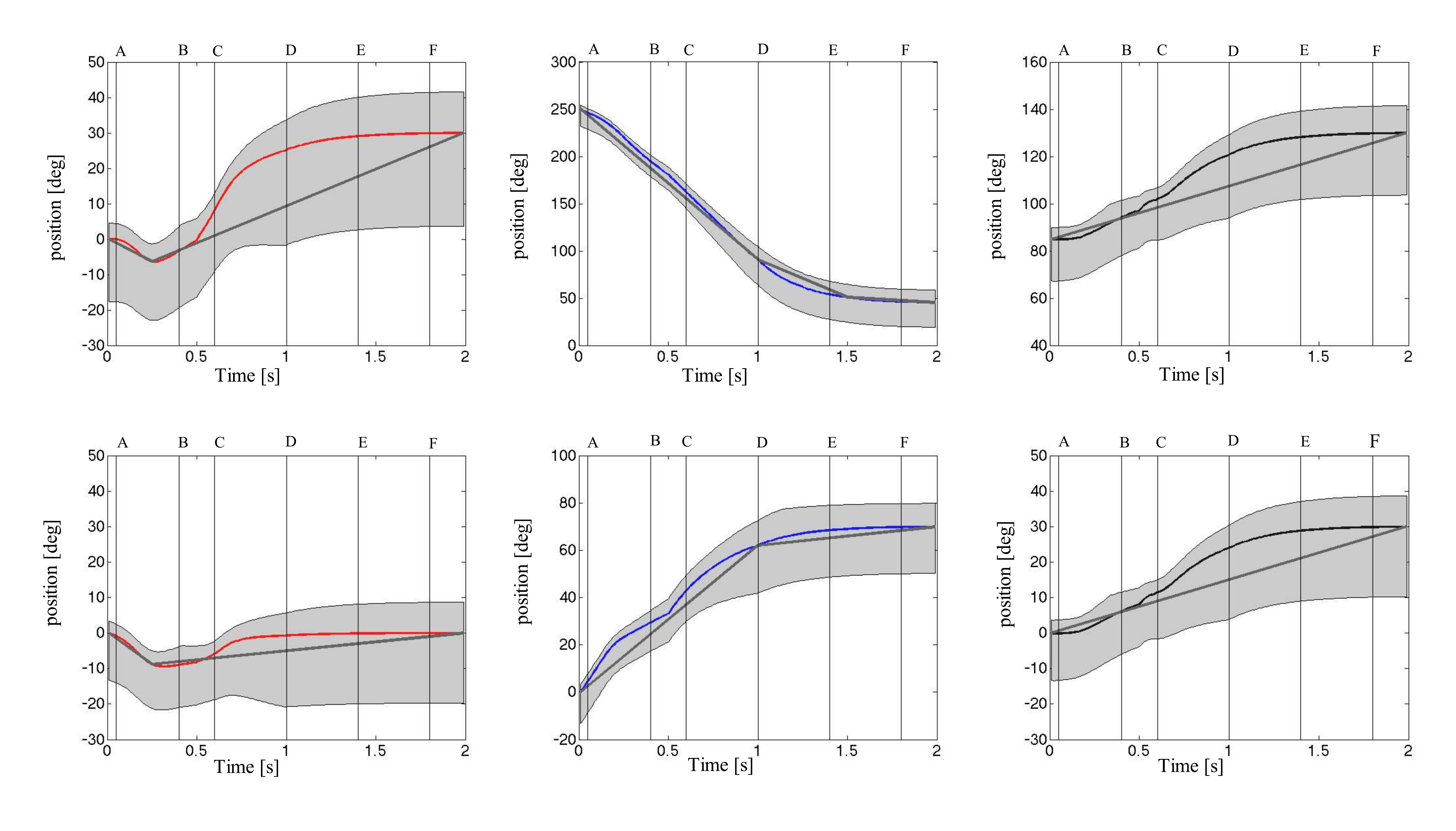}
   \end{tabular}
   \end{center}
   \caption{ \label{Fig:6PositionLinearization} Position profiles that correspond to the pairs of actuators for positioners 1,2, and 3 in Fig.~\ref{Fig:Simulator}. Columns show the position profiles for each positioner and the piecewise linearized estimation of the profiles. The first and second profiles of each column correspond to the first and second actuator of each positioner, respectively. Vertical lines indicate the moment at which the snapshots in Fig.~\ref{Fig:Simulator}  were taken. Each profile of the six motor can be piecewise linearized using maximum two breaking points.}
   \end{figure}

Fig.~\ref{Fig:VelocityProfiles} shows velocity profiles of the both motors of the three positioner shown in the Fig.~\ref{Fig:Simulator}.  Fig.~\ref{Fig:6PositionLinearization} shows the corresponding position profiles to the velocity profiles shown in Fig.~\ref{Fig:VelocityProfiles}. By continuing the steps of the proposed discretization algorithm, we will get piecewise linearized profiles for the motor positions. In such a challenging scenarios with respect to the collision avoidance and discretization, profiles could be linearized with maximum three segments. Representing each velocity profile with a maximum three segments with constant value and yet ensuring collision avoidance is a promising step toward realization of the positioners.  

%%%%%%%%%%%%%%%%%%%%%%%%%%%%%%%%%%%%%%%%%%%%%%%%%%%%%%%%%%%%%
\section{Conclusion}

Robot positioning of a large number of parallel spectrograph fibers such as the ones envisioned for the DESI (5000 positioner robots) or PFS (2400 positioners)
projects are a promising approach for providing a 3D map of large portions of our distant universe. Such dense packing of mechanical robot positioners will result in workspace overlap between neighboring robots. One of the key challenges is
designing optimized trajectory profiles to minimize collision deadlocks.
In this contribution we propose a decentralized method for coordination of the positioners by means of a potential field that guarantees collision avoidance while securing convergence to the target points. 
Simulation results show the feasibility of this method for mid-scale and large-scale spectrograph fiber positioners, with active motion duration of only 2.5 seconds for any number of positioners, under realistic assumptions. An iterative method for piecewise
linearizing of the position profiles of the motors is proposed. The velocity profiles are designed with constant velocity segments.

The main advantage of this method is that the motion time is kept short. This translates directly into longer available observation time and more reliable scientific data. Future research directions will include optimization of the discretization process for further minimization of computational efforts.
%%%%%%%%%%%%%%%%%%%%%%%%%%%%%%%%%%%%%%%%%%%%%%%%%%%%%%%%%%%%%

\acknowledgments     %>>>> equivalent to \section*{ACKNOWLEDGMENTS}       
 
 LM, JPK and LJ acknowledge the support from the European Research Council (ERC) advanced grant "Light on the Dark" (LIDA).
      FP and JS thanks the support from the Spanish MINECO grants AYA10-21231 and MultiDark-CSD2009-0064. 

%%%%%%%%%%%%%%%%%%%%%%%%%%%%%%%%%%%%%%%%%%%%%%%%%%%%%%%%%%%%%
%%%%% References %%%%%

\bibliography{article}   %>>>> bibliography data in report.bib

\begin{thebibliography}{10}

\bibitem{anderson2012clustering}
{\sc Anderson, L., Aubourg, E., Bailey, S., Bizyaev, D., Blanton, M., Bolton,
  A.~S., Brinkmann, J., Brownstein, J.~R., Burden, A., Cuesta, A.~J., et~al.}
\newblock The clustering of galaxies in the sdss-iii baryon oscillation
  spectroscopic survey: baryon acoustic oscillations in the data release 9
  spectroscopic galaxy sample.
\newblock {\em Monthly Notices of the Royal Astronomical Society 427}, 4
  (2012), 3435--3467.

\bibitem{colless20012df}
{\sc Colless, M., Dalton, G., Maddox, S., Sutherland, W., Norberg, P., Cole,
  S., Bland-Hawthorn, J., Bridges, T., Cannon, R., Collins, C., et~al.}
\newblock The 2df galaxy redshift survey: spectra and redshifts.
\newblock {\em Monthly Notices of the Royal Astronomical Society 328}, 4
  (2001), 1039--1063.

\bibitem{de2006formation}
{\sc De~Gennaro, M.~C., and Jadbabaie, A.}
\newblock Formation control for a cooperative multi-agent system using
  decentralized navigation functions.
\newblock In {\em American Control Conference, 2006\/} (2006), IEEE, pp.~6--pp.

\bibitem{dimarogonas2005decentralized}
{\sc Dimarogonas, D.~V., and Kyriakopoulos, K.~J.}
\newblock Decentralized motion control of multiple agents with double
  integrator dynamics.
\newblock In {\em 16th IFAC World Congress, to appear\/} (2005).

\bibitem{ge2012decentralized}
{\sc Ge, F., Wei, Z., Lu, Y., Tian, Y., and Li, L.}
\newblock Decentralized coordination of autonomous swarms inspired by chaotic
  behavior of ants.
\newblock {\em Nonlinear Dynamics 70}, 1 (2012), 571--584.

\bibitem{huchra1983survey}
{\sc Huchra, J., Davis, M., Latham, D., and Tonry, J.}
\newblock A survey of galaxy redshifts. iv-the data.
\newblock {\em The Astrophysical Journal Supplement Series 52\/} (1983),
  89--119.

\bibitem{makarem2011decentralized}
{\sc Makarem, L., and Gillet, D.}
\newblock Decentralized coordination of autonomous vehicles at intersections.
\newblock In {\em World Congress\/} (2011), vol.~18, pp.~13046--13051.

\bibitem{makarem2012fluent}
{\sc Makarem, L., and Gillet, D.}
\newblock Fluent coordination of autonomous vehicles at intersections.
\newblock In {\em Systems, Man, and Cybernetics (SMC), 2012 IEEE International
  Conference on\/} (2012), IEEE, pp.~2557--2562.

\bibitem{makarem2013collision}
{\sc Makarem, L., Kneib, J.-P., Gillet, D., Bleuler, H., Bouri, M., Jenni, L.,
  Prada, F., and Sanchez, J.}
\newblock Collision avoidance in next-generation fiber positioner robotic
  system for large survey spectrograph.
\newblock {\em arXiv preprint arXiv:1312.1644\/} (2013).

\bibitem{rantzer2000piecewise}
{\sc Rantzer, A., and Johansson, M.}
\newblock Piecewise linear quadratic optimal control.
\newblock {\em Automatic Control, IEEE Transactions on 45}, 4 (2000), 629--637.

\bibitem{rewienski2003trajectory}
{\sc Rewienski, M., and White, J.}
\newblock A trajectory piecewise-linear approach to model order reduction and
  fast simulation of nonlinear circuits and micromachined devices.
\newblock {\em Computer-Aided Design of Integrated Circuits and Systems, IEEE
  Transactions on 22}, 2 (2003), 155--170.

\bibitem{sontag1981nonlinear}
{\sc Sontag, E.}
\newblock Nonlinear regulation: The piecewise linear approach.
\newblock {\em Automatic Control, IEEE Transactions on 26}, 2 (1981), 346--358.

\bibitem{tanner2005towards}
{\sc Tanner, H.~G., and Kumar, A.}
\newblock Towards decentralization of multi-robot navigation functions.
\newblock In {\em Robotics and Automation, 2005. ICRA 2005. Proceedings of the
  2005 IEEE International Conference on\/} (2005), IEEE, pp.~4132--4137.

\bibitem{york2000sloan}
{\sc York, D.~G., Adelman, J., Anderson~Jr, J.~E., Anderson, S.~F., Annis, J.,
  Bahcall, N.~A., Bakken, J., Barkhouser, R., Bastian, S., Berman, E., et~al.}
\newblock The sloan digital sky survey: Technical summary.
\newblock {\em The Astronomical Journal 120}, 3 (2000), 1579.

\end{thebibliography}
\bibliographystyle{acm}   %>>>> makes bibtex use spiebib.bst

\end{document}